\magnification=1200
\baselineskip=18truept
\input epsf


\def\draftversion{N}

\if \draftversion Y


\fi

\def\npblong{1}
\def\plbfirst{2}
\def\daemi{3}
\def\kap{4}
\def\soniblum{5}
\def\shamirfur{6}
\def\yoshn{7}
\def\minfty{8}
\def\412{9}
\def\num412{10}
\def\rebbi{11}
\def\yamada{12}
\def\bcasher{13}
\def\ursn{14}
\def\aoki{15}
\def\chem{16}
\def\realsqrt{17}
\def\zenkin{18}
\def\narvra{19}
\def\staiputin{20}
\def\vranas{21}

\line{\hfill RU--97--63}
\centerline {\bf Exactly massless quarks on the lattice.}
\vskip 1cm
\centerline{Herbert Neuberger}
\vskip 1.5cm
\centerline{\it Department of Physics and Astronomy}
\centerline{\it Rutgers University}
\centerline{\it Piscataway, NJ 08855--0849}
\vskip 2cm
\centerline{\bf Abstract}
It is suggested that the fermion determinant for a 
vector-like gauge theory with strictly massless 
quarks can be represented on the lattice as 
$\det{{1+V}\over 2}$, where $V=X(X^\dagger X)^{-1/2}$
and $X$ is the Wilson-Dirac lattice operator with a negative
mass term. There is no undesired doubling and  
no need for any fine tuning.
Several other appealing features of the formula are pointed
out.
\vskip .3cm
\vfill
\eject

From the start, it was awkward to ensure masslessness
of quarks in lattice QCD. Without masslessness there are
no chiral symmetries and no massless pions. Massless QCD
is a beautiful parameter free theory in the continuum, just
like pure YM is. However, while pure YM on the lattice is 
elegant, the addition of quarks makes the lattice theory
less appealing. The overlap [\npblong, \plbfirst,
\daemi] provided a solution to
this state of affairs; however, the formalism is a bit complicated,
and its elegance becomes a matter of personal taste. 
More importantly, the computational cost seems to be
daunting. Domain
walls [\kap, \soniblum, \shamirfur]
provide an approximation to the overlap, but they
sacrifice strict masslessness, so we are back to the more
murky situation of standard formulations. Staggered fermions
do preserve masslessness, but it is a nuisance to have so many
of them, and the connection to continuum operators is messy.
My purpose here is to present a simple formula for the
fermion determinant which I argue is equivalent to the overlap 
but is much more attractive in appearance. Only future work
can tell with any confidence
whether real practical progress has been made, but there are
some very attractive features that I wish to point out.

The basic observation stems from recent work [\yoshn]
on the overlap
in odd dimensions. For $d=2k+1$ the overlap for Dirac fermions
can be written as the determinant of a finite matrix of fixed shape.
This simpler formula descends by dimensional 
reduction from $d+1$ dimensions where it represented Weyl fermions
in a less explicit formulation.
Dimensionally reducing again, a simple formula
for Dirac fermions in $2k$ dimensions is obtained. To make
this letter relatively self-contained, 
I present a more direct derivation, 
involving no dimensional reductions. 

In [\npblong] the overlap for a 
vector-like theory is constructed as follows:
The chiral 
determinant is replaced at the regulated level by
the overlap of two many-body states. 
These are the ground states
of two bilinear Hamiltonians,
$$
{\cal H}^\pm = a^\dagger H^\pm a, \eqno{(1)}$$
with all indices suppressed. The matrices $H^\pm$ are obtained from
$$
H(m) = \pmatrix {B +m & C \cr 
C^\dagger & -B -m \cr}\eqno{(2)}$$
with $H^+ = H(\infty )$, $H^- = H(-m_0 )$ and $0 < m_0 < 2$.
The infinite argument for $H^+$ can be replaced by any finite positive
number, but the equations are somewhat simpler with our choice
[\minfty, \412, \num412]. The matrices $C$ and $B$ are given below:
$$
\eqalign{
( C )_{x \alpha i, y \beta j}
& ={1\over 2} \sum_{\mu=1}^{2k} \sigma_\mu^{\alpha\beta}
[\delta_{y,x+\hat\mu} (U_\mu (x) )_{ij} -
\delta_{x,y+\hat\mu} (U_\mu^\dagger (y))_{ij}] ,\cr
( B )_{x \alpha i, y \beta j} & = {1\over 2} \delta_{\alpha\beta} 
\sum_{\mu =1}^{2k} [2\delta_{xy}
\delta_{ij} - \delta_{y,x+\hat\mu} (U_\mu (x) )_{ij} -
\delta_{x,y+\hat\mu} (U_\mu^\dagger (y))_{ij}].\cr} \eqno{(3)}$$
$x,y$ are sites on the lattice, $\alpha ,\beta$ are Weyl spinor
indices and $i,j$ are color indices. The $\sigma_\mu$ are Euclidean
Weyl matrices in $2k$ dimensions. The overlap $O$, is given by
$$
O=|<v_+ | v_- >|^2,~~~~~~~~~~~
{\cal H}^\pm |v_\pm > = E_{\rm min}^\pm |v_\pm>.
\eqno{(4)}$$
An equivalent representation is also given in [\npblong]:
$$
O=|<v_+ | v_- >||<t_+|t_->|,~~~~~~~~~~~
{\cal H}^\pm |t_\pm > = E_{\rm max}^\pm |t_\pm>.
\eqno{(5)}$$
$E^\pm_{\rm min(max)}$ denote minimal (maximal) energies which
define the associated states, assuming no degeneracies.

I now replace the two sets of fermion operators, one
for each factor in equation (5), by a single
set of double their size:
$$
{\cal H}_2^\pm = A^\dagger H_2^\pm A,~~~~~~~~H_2^\pm =\pmatrix
{0 & H^\pm \cr H^\pm & 0}. \eqno{(6)}$$ Clearly,
$$
O=|<V_+ | V_->|,~~~~~~~~~~~~~
{\cal H}_2^\pm |V_\pm > = E_{\rm min}^\pm |V_\pm>. \eqno{(7)}$$
But, using [\npblong] again, we immediately
can write down the overlap as
$$
O=|\det {{1+\Gamma_{2k+1} \epsilon (H^- ) }\over 2}|,~~~~~~~
\Gamma_{2k+1} =\pmatrix {1 & 0 \cr 0 & -1 },~~
\epsilon (H) \equiv {H\over \sqrt{H^2 }},\eqno{(8)}$$
where we assume $\det H \ne 0$ (this is equivalent
to the absence of degeneracies 
mentioned after equation (5) above). The matrix
$$
V=\Gamma_{2k+1} \epsilon (H^- )~~~~~~~~~(\Gamma_{2k+1}^2=1,
\epsilon (H^- )^2 =1 )\eqno{(9)}$$
is unitary while both factors are hermitian. 
Actually, the absolute sign in eq. (8) is not needed:
$$
\det (1+V) = \prod_{r=1}^{R_0} (1+\lambda_r )
\prod_{c=1}^{C_0} |1+\lambda_c|^2 \ge 0 .
\eqno{(10)}$$
Here, $\lambda_r =\lambda_r^*$ and $\lambda_c\ne\lambda_c^*$. 
Equation (8), without the absolute value signs, 
is the main result of this letter. Noting that 
$$\Gamma_{2k+1} H(m) \equiv X(m) = \pmatrix {B+m & C\cr
-C^\dagger & B+m },\eqno{(11)}$$
and that 
$$\pmatrix {0 & \sigma_\mu \cr \sigma_\mu^\dagger & 0} =\gamma_\mu , 
\eqno{(12)}$$
where the $\gamma_\mu$ are Euclidean Dirac matrices, we obtain
the result announced in the abstract. $X(m)$ is the familiar
Wilson Dirac operator, and we denote $X(-m_0)$ by $X$.

The inverse of the matrix whose determinant we have to take
is easily seen to have only a single massless Dirac particle
pole, so there is no unnecessary doubling [\yoshn]. In some
sense our formula is a realization of 
an idea of Rebbi's [\rebbi], only
some of the ingredients are different and there is a body
of overlap work to rely on [\npblong, \plbfirst, \daemi, \yamada].

One of the most salient properties of strictly massless fermions
are the exact and robust zeros in instanton backgrounds. From
the equivalence to the overlap we know that $V$ should 
have $-1$ as an eigenvalue with 
degeneracy ${1\over 2}|{\rm tr} \epsilon (H^- )|$
(the overlap lattice topological charge is given by
$n_{\rm top} = {1\over 2} {\rm tr} \epsilon (H^- )$) 
and that this property is robust under small 
variations of the background. It is amusing to see this directly in the
particular case of a single instanton or anti-instanton, 
where $|n_{\rm top}| =1$. The following identities 
are easily established:
$$\det (1+V) = \det (1+V^{-1} ) ={{\det (1+V)}\over {\det V}}=
\det (V)\det (1+V).\eqno{(13)}$$
If the number of sites is $\Omega$, the dimension of $V$
is $N\equiv R_0 +2C_0=2^k d_R \Omega$, 
where $d_R$ is the range of group indices. 
Clearly, $N$ is even. 
$$
\det (V) = \det (\Gamma_{2k+1})\det(\epsilon (H^- )) =
(-)^{N\over 2} (-)^{{N\over 2} +n_{\rm top}}=-1\eqno{(14)}$$
Inserting $\det(V)=-1$ into equation (13) gives $\det (1+V) =0$, the
expected result.

Once we are willing to interpret ${{1+V}\over 2}$ as the 
Dirac operator, many applications in four
dimensions suggest themselves. In
the continuum, the spectral density of the massless Dirac operator
per unit four-volume is supposed to concentrate at the origin
to reproduce the chiral symmetry order parameter [\bcasher]. Attempts
to show this on the lattice in the past were hindered by 
the eigenvalues not falling on a line in the complex plane,
and it was difficult, in finite volumes, to identify
and remove the effects of global topological charge, 
which is something one should do. 
Now it would be easy: all the eigenvalues of $V$ 
reside on the unit circle. Those strictly at $-1$ reflect
global topology.
Those close to $-1$, but not exactly at $-1$,  
should build up the condensate $<\bar\psi\psi >$.

A related application is the measurement of $f_\pi$ using
a finite-size soft-pion theorem as in [\ursn]. Since chiral
symmetry is exact we do not have to deal with the more complex 
situation created by the presence of 
an explicit symmetry breaking parameter. 
Anyhow, the interpretation of the transition obtained
by tuning the mass parameter in ordinary Wilson-Dirac  
formulations, while away from the continuum limit, is likely
different from a full restoration of chiral symmetries as
first pointed out by Aoki [\aoki]. Therefore, working with
ordinary Wilson fermions and obtaining 
$f_\pi$ via finite size theorems
is promising to be messy.

Since the eigenvalue distribution of $V$
has the potential of behaving smoothly on the lattice
there is a chance to make some progress
on the painful problem of QCD at finite chemical potential [\chem],
at least in the rather interesting situation of 
strictly massless quarks.

In addition to the overlap, several proposals to regulate
{\it chiral} gauge theories maintain exact gauge invariance
of the absolute value of the chiral determinant [\realsqrt,\zenkin]. 
It is typically proposed [\realsqrt]
to take a square root of $|\det (X)|$ 
with a mass parameter finely tuned to Wilson-Aoki criticality.
Clearly, this is not very appealing, since $\det (X)$ changes
sign frequently in that vicinity (it is real) and the absolute
value will give trouble when variations with respect
to the background fields are taken, for example for the purpose
of deriving a Ward identity.
However, using $\sqrt {\det {{1+V}\over 2}}$ avoids this problem
and is identical to the absolute value of the chiral
determinant in the overlap; as a bonus the exact zeros
induced by nontrivial global topology are also maintained. 
Maybe there is a possibility to unify in this way
several different approaches to the problem of regularizing
chiral gauge theories.  

In recent work [\narvra], it was shown that the method of [\npblong]
for using the overlap to count instantons gives results
in perfect agreement with very different methods
[\staiputin], thus providing
much support for both results. 
By varying $m_0$ in the interval $(0,2)$ a flow of
$V$-eigenvalues on the unit circle would be induced. 
The eigenvalue motions on the unit circle have the potential
of providing a pictorial insight via a scale-dependent 
(the scale is $\sim {1\over m_0}$) 
fermionic probe of the gauge background.

Much of what I said needs further study. Some of the
needed future work is analytical:
With the new formulae, 
perturbation theory may be more compact 
and more Feynman-like. Of course, one needs to 
check whether the natural guess
${2\over {1+V}}$ is indeed an appropriate definition of
the fermion propagator, and how it is related to
the continuum. Numerical work in two dimensions
could check the flow patterns and see whether they are
interesting.
Also, comparisons between the pure overlap and its
domain wall approximations [\vranas] might be facilitated
by the new version of the formula. 
And, most importantly, it is an attractive, if difficult, 
challenge to develop an efficient numerical technique to include
$\det {{1+V}\over 2}$ in dynamical simulations in four dimensions,
or even to compute just the inverse ${2\over {1+V}}$. 
While the problem looks hard, it does seem somewhat easier than
when looking at the older form of the overlap in equation (4). 
\smallskip 
This work was supported in part by the DOE under grant \#
DE-FG05-96ER40559.

\bigskip
\centerline{\bf References.}
\medskip
\item{[\npblong]} R. Narayanan, H. 
Neuberger, Nucl. Phys. B 443 (1995) 305.
\item{[\plbfirst]} R. Narayanan, H. 
Neuberger, Phys. Lett. B 302 (1993) 62.
\item{[\daemi]} S. Randjbar-Daemi and J. Strathdee, 
Phys. Lett. B348 (1995) 543; Nucl. Phys. B443 (1995) 386; 
Phys. Lett. B402 (1997) 134; Nucl. Phys. B461 (1996) 305;
Nucl. Phys. B466 (1996) 335. 
\item{[\kap]} D. B. Kaplan, Phys. Lett. B288 (1992) 342.
\item{[\soniblum]} T. Blum, A. Soni, Phys. Rev. D56 (1997) 174.
\item{[\shamirfur]} V. Furman, Y. Shamir, Nucl. Phys. B439 (1995) 54.
\item{[\yoshn]} Y. Kikukawa, H. Neuberger, hep-lat/9707016.
\item{[\minfty]} D. Boyanovsky, E. Dagotto, E. Fradkin, Nucl. Phys. B 285 
(1987) 340; Y. Shamir, Nucl. Phys. B406 (1993) 90.
\item{[\412]} R. Narayanan, H. Neuberger, Phys. Lett. B 393 (1997) 360;
Y. Kikukawa, R. Narayanan, H. Neuberger, Phys. Lett. B 399 (1997) 105.
\item{[\num412]} Y. Kikukawa, R. Narayanan, H. Neuberger, 
hep-lat/9705006 (1997).
\item{[\rebbi]} C. Rebbi, Phys. Lett. 186B (1987) 200.
\item{[\yamada]} A. Yamada, hep-lat/9705040.
\item{[\bcasher]} T. Banks, A. Casher, Nucl. Phys. B169 (1980) 103.
\item{[\ursn]} U. M. Heller, H. Neuberger, Phys. Lett. 207B (1988) 189.
\item{[\aoki]} S. Aoki, Phys. Rev. D30 (1984) 2653. 
\item{[\chem]} I. M. Barbour, S. E. Morrison, E. G. Klepfish,
J. B. Kogut, M-P Lombardo, hep-lat/9705038.
\item{[\realsqrt]} S. Della Pietra, V. Della Pietra, L. Alvarez-Gaume, 
Commun. Math. Phys. 109 (1987) 691;
K. Okuyama, H. Suzuki, hep-lat/9706163;
G. T. Bodwin, Phys. Rev. D54 (1996) 6497.
\item{[\zenkin]} S. V. Zenkin, Phys. Lett, 395B (1997) 283. 
\item{[\narvra]} R. Narayanan, P. Vranas, hep-lat/9702005.
\item{[\staiputin]} P. de Forcrand, M. Garcia Perez, I-O.
Stamatescu, hep-lat/9701012.
\item{[\vranas]} P. M. Vranas, hep-lat/9705023.

\vfill\eject
\end